# Time Evolution of Density Parameters for Matter and Dark Energy and their Interaction Term in Brans-Dicke Gravity

Sudipto Roy[1], Dipika Nandi[2a], Sumana Ghosh[2b] and Apashanka Das[2c]

[1]Assistant Professor, Department of Physics, St. Xavier's College, Kolkata, India
[2]Postgraduate Student of Physics (2016-18), St. Xavier's College, Kolkata, India

E-mails: [1]roy.sudipto@sxccal.edu, [2a]dipika.nandi949@gmail.com,
[2b]sumanaghosh284@gmail.com, [2c]saanjaay131@gmail.com

## Abstract

In the framework of Brans-Dicke (BD) theory, the first part of the present study determines the time dependence of BD parameter, energy density and equation of state (EoS) parameter of the cosmic fluid in a universe expanding with acceleration, preceded by a phase of deceleration. For this purpose, a scale factor has been chosen such that the deceleration parameter, obtained from it, shows a signature flip with time. Considering the dark energy to be responsible for the entire pressure, the time evolution of energy parameters for matter and dark energy and the EoS parameter for dark energy have been determined. An effective interaction term, between matter and dark energy, has been proposed and calculated. Its negative value at the present time indicates conversion of matter into dark energy. Using this term, the time dependence of the rates of change of matter and dark energy has been determined. It is found that the nature of dependence of the scalar field upon the scale factor plays a very important role in governing the time evolution of the cosmological quantities studied here. The present study provides us with a simple way to determine the time evolution of dark energy for a homogeneous and isotropic universe of zero spatial curvature, without involving any self-interaction potential or cosmological constant in the formulation.

## 1   Introduction

The accelerated expansion of the universe is one of the most interesting and important phenomena in the field of cosmology that have been obtained through astrophysical observations in recent times (Riess et al. 1998; Riess et al. 2001; Bennett et al. 2003; Komatsu et al. 2011; Percival et al. 2010). An exotic form of energy, with a negative pressure, has been found to be responsible for the accelerated expansion of the universe. This energy is known as dark energy (DE). In the fields of physics and astronomy, an extensive research is now taking place, throughout the world, on the nature and dynamics of DE. A number of models have been proposed to explain this accelerated expansion of the universe, following a phase of deceleration. In most of the models, DE is represented by cosmological constant Λ (Carroll 2001).

The present article is based on Brans-Dicke (BD) theory of gravitation. The BD theory is characterized by a scalar field ($\phi$) and a dimensionless coupling parameter ($\omega$) that govern the dynamics of space-time geometry. It can be regarded as a natural extension of the general theory of relativity which is obtained in the limit of an infinite $\omega$ and a constant value of $\phi$ (Sahoo and Singh 2002). The BD theory of gravity can be regarded as one of the most important theories, among all prevalent alternative theories of gravitation, which have very successfully explained the early and late time behaviours of the universe and solved the problems of inflation (Linde 1990). As an extension of the original BD theory, a generalized version was proposed, where $\omega$ is regarded as a function of the scalar field $\phi$ (Bergmann 1968; Wagoner 1970; Nordtvedt Jr. 1970). Several models regarding the expanding universe have been formulated on the basis of this theory (Sahoo and Singh 2002; Banerjee and Ganguly 2009; Chakraborty and Debnath 2009; Jamil and Momeni 2011; Satish and Venkateswarlu 2014).

In the present study, BD field equations (for a spatially flat, homogeneous and isotropic space-time) have been used to determine time dependence of the coupling parameter ($\omega$), equation of state (EoS) parameter ($\gamma$) and energy density ($\rho$). Time dependence of $\omega$ was previously studied by many groups, using scale factors that have power law dependence upon time (Sahoo and Singh 2002; Jamil and Momeni 2011; Satish and Venkateswarlu 2014). These scale factors lead to time independent deceleration parameters that are not consistent with the interpretations of recent astrophysical observations. According to astrophysical observations, the universe has made a transition from a phase of decelerated expansion to a phase of accelerated expansion, implying that the deceleration parameter has a dependence upon time (Riess et al. 2001; Chand, Mishra and Pradhan 2016). Taking this fact into consideration, we have used a scale factor ($a$) that leads to a time dependent deceleration parameter which changes sign from positive to its present negative value. For simplicity, we have assumed power law dependence of $\omega$ upon $\phi$ and also of $\phi$ upon $a$. Time dependence of $\rho$ and the EoS parameter ($\gamma$), in the present study, are consistent with the results obtained from some recent studies based on general relativity with anisotropic space-time (Pradhan, Amirhashchi and Saha 2011; Yadav, Rahaman and Ray 2011). Two theoretical models have been proposed to determine the time variation of density parameters ($\Omega$) for matter and dark energy. The results from this model are in agreement with those obtained from studies based on completely different premises (Das and Al Mamon 2014; Paul, Thakur and Ghose 2010). An effective interaction term ($Q_{eff}$) between matter and dark energy has been calculated here and its time dependence has been studied. Its time variation is found to be consistent with the findings of some other studies (Zadeh, Sheykhi and Moradpour 2017; Cueva and Nucamendi 2010). In terms of $Q_{eff}$ we have determined the time rate of change of densities for matter and dark energy.

## 2 Metric and Field Equations

The action, in the generalized Brans-Dicke theory, is expressed as,

$$\mathcal{A} = \frac{1}{16\pi} \int d^4x \sqrt{-g} \left( \phi R + \frac{\omega(\phi)}{\phi} g^{\mu\nu} \partial_\mu \phi \partial_\nu \phi + \mathcal{L}_m \right) \tag{1}$$

In equation (1), $g$ is the determinant of the metric tensor $g^{\mu\nu}$, $L_m$ is the Lagrangian for matter, $R$ denotes the Ricci scalar, $\varphi$ is the Brans-Dicke scalar field and $\omega$ is a dimensionless coupling parameter. This parameter is regarded as a function of $\varphi$ in generalized BD theory. The variation of action in equation (1) leads to the following field equations.

$$R_{\mu\nu} - \frac{1}{2} R g_{\mu\nu} = \frac{1}{\phi} T_{\mu\nu} - \frac{\omega}{\phi^2} \left[ \phi_{,\mu} \phi_{,\nu} - \frac{1}{2} g_{\mu\nu} \phi_{,\beta} \phi^{,\beta} \right] - \frac{1}{\phi} \left[ \phi_{\mu;\nu} - g_{\mu\nu} \left\{ \frac{\partial^2 \phi}{\partial t^2} - \nabla^2 \phi \right\} \right] \tag{2}$$

$$\frac{\partial^2 \phi}{\partial t^2} - \nabla^2 \phi = \frac{1}{2\omega+3} T \tag{3}$$

$R_{\mu\nu}$ is the Ricci tensor and $T_{\mu\nu}$ is the energy-momentum tensor. In the above equations, a semicolon stands for a covariant derivative and a comma denotes an ordinary derivative with respect to $x^\beta$.

The energy-momentum tensor ($T_{\mu\nu}$) for the cosmic constituents is given by,

$$T_{\mu\nu} = (\rho + p) u_\mu u_\nu + g_{\mu\nu} p \tag{4}$$

Here, $\rho$ denotes the energy density, $p$ is the isotropic pressure, $u_\nu$ is the four-velocity vector and $T$ is the trace of $T_j^i$. This form of the energy-momentum tensor is based on an assumption that the total matter-energy content of the expanding universe is a perfect fluid. In a co-moving coordinate system, $u^\nu = (0,0,0,1)$, having the characteristic of $g_{\mu\nu} u^\mu u^\nu = 1$.

The line element for a homogeneous and isotropic universe, in Friedmann-Robertson-Walker (FRW) cosmology, can be expressed as,

$$ds^2 = -dt^2 + a^2(t) \left[ \frac{dr^2}{1-kr^2} + r^2 d\theta^2 + r^2 \sin^2\theta d\xi^2 \right] \tag{5}$$

Where $a(t)$ denotes the scale factor, $t$ is the cosmic time, $k$ is the spatial curvature parameter. Three spherical polar coordinates are denoted by $r$, $\theta$ and $\xi$ respectively. Here $k$ describes the closed, flat and open universes corresponding to $k = 1, 0, -1$ respectively.

The following three equations are the field equations of BD theory for a universe filled with a perfect fluid, in the geometry of FRW space-time expressed by equation (5). These equations have been obtained from the field equations (2) and (3) in which $T_{\mu\nu}$ from equation (4) has been incorporated and the components of the metric tensor $g_{\mu\nu}$ have been taken from the line element expressed by equation (5).

$$3\frac{\dot{a}^2+k}{a^2} + 3\frac{\dot{a}}{a}\frac{\dot{\phi}}{\phi} - \frac{\omega(\phi)}{2}\frac{\dot{\phi}^2}{\phi^2} = \frac{\rho}{\phi} \tag{6}$$

$$2\frac{\ddot{a}}{a} + \frac{\dot{a}^2+k}{a^2} + \frac{\omega(\phi)}{2}\frac{\dot{\phi}^2}{\phi^2} + 2\frac{\dot{a}}{a}\frac{\dot{\phi}}{\phi} + \frac{\ddot{\phi}}{\phi} = -\frac{\gamma\rho}{\phi} \tag{7}$$

$$\frac{\ddot{\phi}}{\phi} + 3\frac{\dot{a}\dot{\phi}}{a\phi} = \frac{\rho}{\phi}\frac{(1-3\gamma)}{2\omega+3} - \frac{\dot{\phi}}{\phi}\frac{\dot{\omega}}{2\omega+3} \tag{8}$$

In these field equations, $\gamma (\equiv P/\rho)$ is the equation of state (EoS) parameter for the cosmic fluid, which has been treated as a function of time in the present study.

## 3 Theoretical Model

Combining equations (6), (7) and (8), for zero spatial curvature ($k = 0$), one obtains,

$$\dot{\omega} + \left(2\frac{\ddot{\phi}}{\dot{\phi}} + 6\frac{\dot{a}}{a} - \frac{\dot{\phi}}{\phi}\right)\omega - 6\left(\frac{\dot{a}^2}{a^2} + \frac{\ddot{a}}{a}\right)\frac{\phi}{\dot{\phi}} = 0 \tag{9}$$

In the present study we have assumed the following ansatzes for $\phi$ and $\omega$. These parameters have been assumed to have power-law relations with $a$ and $\varphi$ respectively.

$$\phi = \phi_0 \left(\frac{a}{a_0}\right)^n \qquad (10)$$

$$\omega = \omega_0 \left(\frac{\phi}{\phi_0}\right)^m \qquad (11)$$

The expression of $\phi$ in equation (10) has been taken from some earlier studies in this regard (Banerjee and Ganguly 2009; Chand, Mishra and Pradhan 2016). The reason for choosing the above empirical relation for $\omega$ is the fact that this is regarded as a function of the scalar field in the generalized Brans-Dicke theory (Wagoner 1970; Nordtvedt Jr. 1970). Combining equation (10) with (11), one gets,

$$\omega = \omega_0 \left(\frac{a}{a_0}\right)^{mn} \qquad (12)$$

Using the expressions (10) and (11) in equation (9), one obtains,

$$m\omega n + (n + 4 - 2q)\omega - \frac{6}{n}(1 - q) = 0 \qquad (13)$$

Here, $q \ (\equiv -\ddot{a}a/\dot{a}^2)$ is the deceleration parameter.
Writing $\omega = \omega_0$ and $q = q_0$ (i.e., their values at $t = t_0$) in equation (13), we get,

$$m = \frac{6(1-q_0)}{n^2 \omega_0} - \frac{n+4-2q_0}{n} \qquad (14)$$

Combining equation (6) with (10) and taking $k = 0$ (for flat space), one gets,

$$\rho = \phi H^2 \left(3 + 3n - \frac{\omega}{2}n^2\right) \qquad (15)$$

Here, $H \ (\equiv \dot{a}/a)$ is the Hubble parameter.
Replacing all parameters in equation (15) by their values at $t = t_0$, one obtains,

$$\omega_0 = \frac{2}{n^2 H_0^2}\left(3H_0^2 + 3nH_0^2 - \rho_0/\phi_0\right) \qquad (16)$$

Using equation (16) in equation (14) one gets,

$$m = \frac{3(1-q_0)H_0^2}{3H_0^2 + 3nH_0^2 - \rho_0/\phi_0} - \frac{n+4-2q_0}{n} \qquad (17)$$

Equations (16) and (17) show that both $\omega_0$ and $m$ depends upon the parameter $n$, which determines the time dependence of $\phi$, as per equation (10).
Using equations (10) and (15) in (7), for $k = 0$ (zero spatial curvature) and writing $\frac{\ddot{a}}{a} = -qH^2$ one obtains the following expression of the equation of state parameter.

$$\gamma = \frac{2q - 1 - 0.5\omega n^2 - n - n^2 + nq}{3 + 3n - 0.5\omega n^2} \qquad (18)$$

The value of the equation of state parameter at the present time is therefore given by,

$$\gamma_0 = \frac{2q_0 - 1 - 0.5\omega_0 n^2 - n - n^2 + nq_0}{3 + 3n - 0.5\omega_0 n^2} \qquad (19)$$

To use the expression of $\omega$ (eqn. 11), the values of $\omega_0$ and $m$ should be taken from equations (16) and (17) respectively. The value of $\omega$, required for the expressions of $\rho$ and $\gamma$, in equations (15) and (18) respectively, should then be taken from equation (11). The value of $\phi$, required for equation (15), should be taken from equation (10).

To determine the time dependence of several cosmological parameters ($\phi, \omega, \rho, \gamma$), following empirical expression of the scale factor has been used in the present model.

$$a = a_0 Exp[\alpha\{(t/t_0)^\beta - 1\}] \qquad (20)$$

This scale factor has been so chosen that it generates a deceleration parameter ($q = -\ddot{a}a/\dot{a}^2$) which changes sign from positive to negative, as a function of time. This signature flip indicates a transition of the universe from a phase of decelerated expansion to a phase of accelerated expansion, in accordance with several recent studies based on astrophysical observations (Riess et al. 1998; Riess et al. 2001). We have taken $a_0 = 1$ for all calculations. Here $\alpha, \beta$ should have the same sign to ensure an increase of the scale factor with time. Using equation (20), the Hubble parameter ($H$) and the deceleration parameter ($q$) are obtained as,

$$H = \frac{\dot{a}}{a} = \frac{\alpha\beta}{t_0}\left(\frac{t}{t_0}\right)^{\beta-1} \qquad (21)$$

$$q = -\frac{\ddot{a}a}{\dot{a}^2} = -1 + \frac{1-\beta}{\alpha\beta}\left(\frac{t}{t_0}\right)^{-\beta} \qquad (22)$$

For $0 < \beta < 1$ and $\alpha > 0$ one finds that, $q \to +\infty$ as $t \to 0$ and $q \to -1$ as $t \to \infty$, showing clearly a signature flip of $q$ with time.

Taking $H = H_0$ and $q = q_0$, at $t = t_0$, the values of the constants $\alpha$ and $\beta$ are obtained as,

$$\alpha = \frac{H_0 t_0}{1 - H_0 t_0 (1 + q_0)} \qquad (23)$$

$$\beta = 1 - H_0 t_0 (1 + q_0) \qquad (24)$$

It is often necessary to find the evolution of a cosmological quantity as a function of the redshift parameter ($z$) where $z = \frac{a_0}{a} - 1$. Using equation (20) one gets the following expression as a relation between redshift parameter and time.

$$z = \frac{a_0}{a} - 1 = Exp[-\alpha\{(t/t_0)^\beta - 1\}] - 1 \qquad (25)$$

Using equations (20-24), one can determine the time dependence of $\phi, \omega, \rho$ and $\gamma$ from (10), (12), (15) and (18) respectively.

The values of different cosmological parameters used in this article are:

$H_0 = \frac{72\frac{Km}{s}}{Mpc} = 2.33 \times 10^{-18} sec^{-1}$, $q_0 = -0.55$, $\rho_0 = 9.9 \times 10^{-27} Kgm^{-3}$, $\Omega_{D0} = 0.7$

$\varphi_0 = \frac{1}{G_0} = 1.498 \times 10^{10} Kgs^2m^{-3}$, $t_0 = 1.4 \times 10^{10} Years = 4.42 \times 10^{17} s$

## 4 Determination of Density Parameters

The total pressure ($P$) of the entire matter-energy content of the universe is contributed by dark energy because, the whole matter content (dark matter + baryonic matter) is regarded as pressureless dust (Banerjee and Ganguly 2009; Farajollahi and Mohamadi 2010). Thus we can write,

$$P = \gamma\rho = \gamma_D \rho_D \tag{26}$$

Here, $\gamma_D$ is the EoS parameter for dark energy and $\rho_D$ is the density of dark energy. Using equation (26), the density parameter for dark energy is given by,

$$\Omega_D = \frac{\rho_D}{\rho} = \frac{\gamma}{\gamma_D} \tag{27}$$

Let us assume the following empirical expression for the EoS parameter of dark energy.

$$\gamma_D = \gamma\, f(t) \tag{28}$$

Here $f(t)$ is a function of time for which we have chosen two empirical forms in the following two models. The reason for assuming the above ansatz is the remarkable similarity of the nature of time evolution of the EoS parameter for dark energy (Vinutha et al. 2016; Pasqua and Chattopadhyay 2016) with that for the total matter-energy content (Pradhan, Amirhashchi and Saha 2011; Yadav, Rahaman and Ray 2011) of the universe. Here $f(t)$ is assumed to be a function that connects $\gamma_D$ to $\gamma$.

### 4.1 Model - 1

We propose the following form of $f(t)$ for model-1 of density parameter calculation.

$$f(t) = A \left(\frac{t}{t_0}\right)^\mu \tag{29}$$

Here $A$ and $\mu$ are dimensionless constants.

Using equations (28) and (29) in (27) we get,

$$\Omega_D = \frac{\rho_D}{\rho} = \frac{\gamma}{\gamma_D} = A^{-1} \left(\frac{t}{t_0}\right)^{-\mu} \tag{30}$$

Taking $\Omega_D = \Omega_{D0}$ at the present time, i.e., at $t = t_0$, in equation (30), we get,

$$A = \frac{1}{\Omega_{D0}} \tag{31}$$

Using equation (31) in (30) we get,

$$\Omega_D = \Omega_{D0} \left(\frac{t}{t_0}\right)^{-\mu} \tag{32}$$

Using equation (32), the density parameter for matter is given by,

$$\Omega_m = \frac{\rho_m}{\rho} = \frac{\rho - \rho_D}{\rho} = 1 - \Omega_D = 1 - \Omega_{D0} \left(\frac{t}{t_0}\right)^{-\mu} \tag{33}$$

In deriving equation (33), we have used the fact that the present universe is composed mainly of matter and dark energy, considering all other forms of energy to be negligibly small (Pal 2000; Goswami 2017).

Using equations (32) and (33), the expressions for densities of dark energy and matter can be respectively written as,

$$\rho_D = \rho\Omega_D = \rho\Omega_{D0}\left(\frac{t}{t_0}\right)^{-\mu} \tag{34}$$

$$\rho_m = \rho\Omega_m = \rho\left[1 - \Omega_{D0}\left(\frac{t}{t_0}\right)^{-\mu}\right] \tag{35}$$

Using equations (29) and (31) in (28), the equation-of-state parameter for dark energy ($\gamma_D$) becomes,

$$\gamma_D = \frac{\gamma}{\Omega_D} = \frac{\gamma}{\Omega_{D0}}\left(\frac{t}{t_0}\right)^{\mu} \tag{36}$$

Using equation (25) in (32) and (33), the expressions of $\Omega_D$ and $\Omega_m$ can be written in terms of redshift ($z$) as,

$$\Omega_D = \Omega_{D0}\left(1 + \frac{1}{\alpha}\ln\frac{1}{z+1}\right)^{-\mu/\beta} \tag{37}$$

$$\Omega_m = 1 - \Omega_{D0}\left(1 + \frac{1}{\alpha}\ln\frac{1}{z+1}\right)^{-\mu/\beta} \tag{38}$$

The present value of $\Omega_{D0}$ is close to 0.7 according to several astrophysical observations (Pal 2000; Goswami 2017). In the evolution of density parameters of the universe, there was a time in the recent past when $\Omega_D = \Omega_m = 0.5$ and, the corresponding z value was lying somewhere in the range of $0 < z < 1$, as obtained from some recent studies (Das and Al Mamon 2014; Paul, Thakur and Ghose 2010) in this regard. That phase of the universe might be same as (or close to) the one when there was a transition from a decelerated expansion to an accelerated expansion of the universe because it is the dark energy that is known to be responsible for the accelerated expansion. According to a recent study, the transition of the universe from a phase of decelerated expansion to its present state of accelerated expansion took place in the past at $z = 0.6818$ which was around $7.2371 \times 10^9$ years ago (Goswami 2017). Choosing $z_c$ to denote the value of the redshift parameter at which the universe had $\Omega_D = \Omega_m$, we get the following expression of $\mu$ from equations (37) and (38).

$$\mu = \frac{\beta \ln(2\Omega_{D0})}{\ln\left[1+\frac{1}{\alpha}\ln\left(\frac{1}{z_c+1}\right)\right]} \tag{39}$$

Using equations (10), (20), (32) and (33), $\Omega_D$ and $\Omega_m$ can be expressed as functions of the scalar field ($\phi$) by the following two equations.

$$\Omega_D = \Omega_{D0}\left(1 + \frac{1}{n\alpha}\ln\frac{\phi}{\phi_0}\right)^{-\mu/\beta} \tag{40}$$

$$\Omega_m = 1 - \Omega_{D0}\left(1 + \frac{1}{n\alpha}\ln\frac{\phi}{\phi_0}\right)^{-\mu/\beta} \tag{41}$$

### 4.2  Model – 2

We propose the following form of $f(t)$ for model-2 of density parameter calculation.

$$f(t) = A\, Exp\left(\lambda \frac{t}{t_0}\right) \tag{42}$$

Where, $A$ and $\lambda$ are dimensionless constants.
Combining equations (27), (28) and (42) we get,

$$\Omega_D = \frac{1}{f(t)} = \frac{1}{A} Exp\left(-\lambda \frac{t}{t_0}\right) \tag{43}$$

Using the fact that $\Omega_D = \Omega_{D0}$ at $t = t_0$ in (43) we get,

$$A = \frac{Exp(-\lambda)}{\Omega_{D0}} \tag{44}$$

Substituting for A in equation (43) from equation (44) we get,

$$\Omega_D = \Omega_{D0} Exp\left\{\lambda\left(1 - \frac{t}{t_0}\right)\right\} \tag{45}$$

Here $\Omega_{D0}$ is approximately equal to 0.7, according to various studies on dark energy (Pal 2000; Goswami 2017).
Using equation (45), the density of dark energy is given by,

$$\rho_D = \rho\Omega_D = \rho\Omega_{D0} Exp\left\{\lambda\left(1 - \frac{t}{t_0}\right)\right\} \tag{46}$$

Using equation (45), the density parameter for matter ( dark + baryonic ) can be expressed as,

$$\Omega_m = \frac{\rho_m}{\rho} = \frac{\rho - \rho_D}{\rho} = 1 - \Omega_D = 1 - \Omega_{D0} Exp\left\{\lambda\left(1 - \frac{t}{t_0}\right)\right\} \tag{47}$$

Using equation (47), the density of matter can be expressed as,

$$\rho_m = \rho\Omega_m = \rho\left[1 - \Omega_{D0} Exp\left\{\lambda\left(1 - \frac{t}{t_0}\right)\right\}\right] \tag{48}$$

Using equations (28), (42) and (44), the equation-of-state parameter for the dark energy ($\gamma_D$) can be expressed as,

$$\gamma_D = \frac{\gamma}{\Omega_{D0}} Exp\left\{-\lambda\left(1 - \frac{t}{t_0}\right)\right\} \tag{49}$$

Combining equations (25), (45) and (47), the expressions of $\Omega_D$ and $\Omega_m$ can be written in terms of redshift ($z$) as,

$$\Omega_D = \Omega_{D0}\, Exp\left[-\lambda\left\{\left[1 - \frac{\ln(z+1)}{\alpha}\right]^{1/\beta} - 1\right\}\right] \tag{50}$$

$$\Omega_m = 1 - \Omega_{D0}\, Exp\left[-\lambda\left\{\left[1 - \frac{\ln(z+1)}{\alpha}\right] - 1\right\}\right] \tag{51}$$

Taking $\Omega_D = \Omega_m$ at $z = z_c$, one gets the following expression of $\lambda$ from equations (50) and (51).

$$\lambda = \frac{ln\left(\frac{1}{2\Omega_{D0}}\right)}{\left[1-\left\{1-\frac{ln(z_c+1)}{\alpha}\right\}^{1/\beta}\right]} \tag{52}$$

Using equations (10) and (20) in (45) and (47), $\Omega_D$ and $\Omega_m$ can be expressed as functions of the scalar field ($\phi$) by the following two equations.

$$\Omega_D = \Omega_{D0} Exp\left\{-\lambda\left(\left[\frac{ln(\phi/\phi_0)}{n\alpha}+1\right]^{1/\beta}-1\right)\right\} \tag{53}$$

$$\Omega_m = 1 - \Omega_{D0} Exp\left\{-\lambda\left(\left[\frac{ln(\phi/\phi_0)}{n\alpha}+1\right]^{1/\beta}-1\right)\right\} \tag{54}$$

### 4.3   A Comparison of Two Models for Density Parameter Calculation

For a positive value of $z_c$, both $\mu$ and $\lambda$ are negative, as obtained from model-1 (eqn. 39) and model-2 (eqn. 52) respectively, indicating an increase of $\Omega_D$ ( and a corresponding decrease of $\Omega_m$) with time. Qualitatively, these two models show the same nature of time dependence for $\Omega_D$ and $\Omega_m$, with $\Omega_D = \Omega_{D0}$ and $\Omega_m = \Omega_{m0}$ at $z = 0$ (present time). The time derivatives of them, for two models, are respectively given by equations (55) and (56).

$$\dot{\Omega}_D = \frac{-\mu}{t_0}\Omega_{D0}\left(\frac{t}{t_0}\right)^{-\mu-1} \tag{55}$$

$$\dot{\Omega}_D = \frac{-\lambda}{t_0}\Omega_{D0} Exp\left\{\lambda\left(1-\frac{t}{t_0}\right)\right\} \tag{56}$$

Since $\mu$ and $\lambda$ are negative, both expressions of $\dot{\Omega}_D$ are positive, being proportional to $\frac{\Omega_{D0}}{t_0}$ at the present time. Near the present time, they have almost the same behaviour (fig. 5).

## 5   Calculation of an Effective Interaction Term

The energy conservation equation is expressed as (Sahoo and Singh 2002),

$$\dot{\rho} + 3H(\rho + P) = 0 \tag{57}$$

Taking the pressure $P = \gamma\rho = \gamma_D\rho_D$ (eqn. 26) and $\rho = \rho_m + \rho_D$ in equation (57) one gets,

$$\dot{\rho}_m + \dot{\rho}_D + 3H[\rho_m + \rho_D(1+\gamma_D)] = 0 \tag{58}$$

If it is assumed that the two entities, matter and dark energy the universe, have been interacting with each other, causing the generation of one of them at the cost of the other, one may define a parameter representing their interaction, on the basis of equation (58). The interaction term ($Q$) can be represented by the following equations (Das and Al Mamon 2014; Zadeh, Sheykhi and Moradpour 2017; Farajollahi and Mohamadi 2010).

$$\dot{\rho}_m + 3H\rho_m = Q \tag{59}$$

$$\dot{\rho}_D + 3H\rho_D(1+\gamma_D) = -Q \tag{60}$$

A negative value of $Q$ represents a transfer of energy from the matter field to the field of dark energy and a positive value of $Q$ represents conversion of dark energy into matter. It is evident from equations (59) and (60) that $Q$ has a dependence upon time and, its values, obtained from these two equations, would not be the same. Let us denote the values of $Q$, obtained from these two equations, by $Q_1$ and $Q_2$ respectively.

Using the relations $\rho = \rho_m + \rho_D$ and $\rho_D = \rho\Omega_D$ in (59) we get,

$$Q_1 = (1 - \Omega_D)\dot{\rho} - \rho\dot{\Omega}_D + 3H\rho(1 - \Omega_D) \tag{61}$$

Using the relations $\rho_D = \rho\Omega_D$ and $\gamma_D = \frac{\gamma}{\Omega_D}$ (eqn. 36) in equation (60) we get,

$$Q_2 = -\Omega_D\dot{\rho} - \rho\dot{\Omega}_D - 3H\rho(\Omega_D + \gamma) \tag{62}$$

To estimate the difference ($\Delta Q$) between $Q_1$ and $Q_2$, we write the following expression.

$$\Delta Q = Q_1 - Q_2 = \dot{\rho} + 3H\rho(1 + \gamma) \tag{63}$$

To have an estimate of an effective interaction between matter and dark energy, we define a parameter $Q_{eff}$ as the average of $Q_1$ and $Q_2$. Thus we have,

$$Q_{eff} = \frac{Q_1 + Q_2}{2} = \left(\frac{1}{2} - \Omega_D\right)\dot{\rho} - \rho\dot{\Omega}_D + \frac{3}{2}H\rho(1 - 2\Omega_D - \gamma) \tag{64}$$

To judge the plausibility of using $Q_{eff}$ as a measure of interaction between matter and dark energy, we have calculated the following expression of $\Delta Q/Q_{eff}$, using equations (63), (64).

$$\frac{\Delta Q}{Q_{eff}} = \frac{\dot{\rho} + 3H\rho(1+\gamma)}{\left(\frac{1}{2}-\Omega_D\right)\dot{\rho} - \rho\dot{\Omega}_D + \frac{3}{2}H\rho(1-2\Omega_D-\gamma)} \tag{65}$$

Here we have used model-1 of density parameter calculation, because two models predict almost the same nature of time dependence of density parameters near the present time (fig. 5). To determine the time dependence of the interaction term ($Q$), we have calculated $\dot{\rho}$ and $\dot{\Omega}_D$, based on the expressions of $\rho$ and $\Omega_D$ in equations (15) and (32) respectively.

$$\dot{\rho} = \rho H\left[n - 2(1+q) - \frac{\omega m n^3}{6 + 6n - \omega n^2}\right] \tag{66}$$

$$\dot{\Omega}_D = \frac{-\mu\Omega_D}{t} = \frac{-\mu\Omega_D}{t_0}\left(\frac{Ht_0}{\alpha\beta}\right)^{\frac{1}{1-\beta}} \tag{67}$$

Equations (10) and (11) have been used, along with the definitions of $H$ and $q$, to calculate $\dot{\rho}$ (eqn. 66) from (15). Equation (21) has been used to calculate $\dot{\Omega}_D$ (eqn. 67) from (32).
Using equations (66) and (67), in (61), (62), (64) and (65) one obtains,

$$Q_1 = (1 - \Omega_D)\rho H\left[n - 2(1+q) - \frac{\omega m n^3}{6+6n-\omega n^2}\right] + \frac{\mu\rho\Omega_D}{t_0}\left(\frac{Ht_0}{\alpha\beta}\right)^{\frac{1}{1-\beta}} + 3H\rho(1 - \Omega_D) \tag{68}$$

$$Q_2 = -\Omega_D\rho H\left[n - 2(1+q) - \frac{\omega m n^3}{6+6n-\omega n^2}\right] + \frac{\mu\rho\Omega_D}{t_0}\left(\frac{Ht_0}{\alpha\beta}\right)^{\frac{1}{1-\beta}} - 3H\rho(\Omega_D + \gamma) \tag{69}$$

$$Q_{eff} = \left(\frac{1}{2} - \Omega_D\right)\rho H\left[n - 2(1+q) - \frac{\omega m n^3}{6+6n-\omega n^2}\right] + \frac{\mu\rho\Omega_D}{t_0}\left(\frac{Ht_0}{\alpha\beta}\right)^{\frac{1}{1-\beta}} + \frac{3}{2}H\rho(1 - 2\Omega_D - \gamma) \quad (70)$$

$$\frac{\Delta Q}{Q_{eff}} = \frac{\rho H\left[n-2(1+q)-\frac{\omega m n^3}{6+6n-\omega n^2}\right]+3H\rho(1+\gamma)}{\left(\frac{1}{2}-\Omega_D\right)\rho H\left[n-2(1+q)-\frac{\omega m n^3}{6+6n-\omega n^2}\right]-\rho\frac{-\mu\Omega_D}{t_0}\left(\frac{Ht_0}{\alpha\beta}\right)^{\frac{1}{1-\beta}}+\frac{3}{2}H\rho(1-2\Omega_D-\gamma)} \quad (71)$$

If the value of $\frac{\Delta Q}{Q_{eff}}$ is found to be sufficiently small, it would be quite reasonable to believe that $Q_{eff}$, defined as an average of $Q_1$ and $Q_2$, will truly represent the interaction between matter and dark energy. Table-1 shows the values of $(Q_{eff})_{t=t_0}$ and $(\Delta Q/Q_{eff})_{t=t_0}$ for several values of the parameter $n$. Here, the present value of $Q_{eff}$ is numerically of the order of $10^{-44}$ and it is negative. The present value of $\Delta Q/Q_{eff}$ is so small ($\sim 10^{-15}$) that $Q_{eff}$, as defined by eqn. 64, is likely to describe the interaction with sufficient accuracy. Its negative value indicates decay of matter into dark energy at the present time. As $n$ becomes more negative, $(Q_{eff})_{t=t_0}$ becomes more negative. It implies that a greater rate of change of the scalar field ($\phi$) has a connection to a larger interaction between mater and dark energy of the universe, causing a larger rate of generation of dark energy at the expense of matter.

| TABLE - 1 | | |
|---|---|---|
| Values of $Q_{eff}$ and $\|\Delta Q/Q_{eff}\|$ at the present time for several values of the parameter $n$ which determines the time variation of the scalar field ($\phi$). | | |
| $n$ | $(Q_{eff})_{t=t_0}$ | $\left\|\left(\frac{\Delta Q}{Q_{eff}}\right)_{t=t_0}\right\|$ |
| -1.60 | $-1.153 \times 10^{-44}$ | $1.296 \times 10^{-14}$ |
| -1.62 | $-1.606 \times 10^{-44}$ | $3.719 \times 10^{-15}$ |
| -1.64 | $-2.046 \times 10^{-44}$ | $2.433 \times 10^{-15}$ |
| -1.66 | $-2.473 \times 10^{-44}$ | $3.020 \times 10^{-15}$ |
| -1.68 | $-2.886 \times 10^{-44}$ | $1.021 \times 10^{-15}$ |
| -1.70 | $-3.285 \times 10^{-44}$ | $9.549 \times 10^{-15}$ |
| -1.72 | $-3.670 \times 10^{-44}$ | $4.883 \times 10^{-15}$ |
| -1.74 | $-4.042 \times 10^{-44}$ | $4.188 \times 10^{-15}$ |
| -1.76 | $-4.400 \times 10^{-44}$ | $2.602 \times 10^{-15}$ |
| -1.78 | $-4.744 \times 10^{-44}$ | $2.001 \times 10^{-15}$ |
| -1.80 | $-5.075 \times 10^{-44}$ | $1.014 \times 10^{-15}$ |
| -1.82 | $-5.392 \times 10^{-44}$ | $6.463 \times 10^{-15}$ |
| -1.84 | $-5.696 \times 10^{-44}$ | $2.797 \times 10^{-15}$ |
| -1.86 | $-5.986 \times 10^{-44}$ | $1.663 \times 10^{-15}$ |

Time dependence of $\dot{\rho}_m$ and $\dot{\rho}_D$ can be calculated from equations (59) and (60) by taking $Q = Q_{eff}$. Use of these two equations for this purpose, instead of (48) and (46), ensures consistency with the energy conservation equation given by (57). Thus we can write,

$$\dot{\rho}_m = Q_{eff} - 3H\rho_m \quad (72)$$

$$\dot{\rho}_D = -Q_{eff} - 3H\rho_D(1 + \gamma_D) \quad (73)$$

Combining equation (70) with (72) and (73), one gets the following expressions (eqns. 74 and 75) for the time derivatives of matter and dark energy ($\dot{\rho}_m$ and $\dot{\rho}_D$).

$$\dot{\rho}_m = \left(\frac{1}{2} - \Omega_D\right)\rho H \left[n - 2(1+q) - \frac{\omega m n^3}{6 + 6n - \omega n^2}\right] + \frac{\mu \rho \Omega_D}{t_0}\left(\frac{Ht_0}{\alpha\beta}\right)^{\frac{1}{1-\beta}}$$
$$+ \frac{3}{2}H\rho(1 - 2\Omega_D - \gamma) - 3H\rho(1 - \Omega_D) \tag{74}$$

$$\dot{\rho}_D = -\left(\frac{1}{2} - \Omega_D\right)\rho H \left[n - 2(1+q) - \frac{\omega m n^3}{6 + 6n - \omega n^2}\right] - \frac{\mu \rho \Omega_D}{t_0}\left(\frac{Ht_0}{\alpha\beta}\right)^{\frac{1}{1-\beta}}$$
$$- \frac{3}{2}H\rho(1 - 2\Omega_D - \gamma) - 3H\rho\Omega_D(1 + \gamma_D) \tag{75}$$

Using these relations we have studied the time dependence of $\dot{\rho}_m$ and $\dot{\rho}_D$ graphically (figs. 13-16). It is evident here that the nature of their time variation must depend upon the parameters $n$ and $\mu$, where $\mu$ is connected to the value of $z_c$, as given by equation (39). A negative value of $\dot{\rho}_m$ and a simultaneous positive value of $\dot{\rho}_D$ imply a conversion of matter into dark energy.

## 6   Choice of Values for the Parameter *n*

The parameter $n$ controls the time evolution of the scalar field ($\phi$). Time variations of $\omega$, $\rho$ and $\gamma$ depend on $n$, as we know that both $m$ and $\omega_0$ are functions of $n$. The scalar field ($\phi$) is the reciprocal of the gravitational constant ($G$) (Banerjee and Ganguly 2009; Roy, Chattopadhyay and Pasqua 2013). Hence, using equations (10) and (20), we can write,

$$G = \frac{1}{\phi} = \frac{1}{\phi_0}\left(\frac{a}{a_0}\right)^{-n} = \frac{1}{\phi_0}Exp[-n\alpha\{(t/t_0)^\beta - 1\}] \tag{76}$$

According to some studies, the gravitational constant increases with time (Pradhan, Saha and Rikhvitsky 2015; Saha, Rikhvitsky and Pradhan 2015; Mukhopadhyay, Chakraborty, Ray and Usmani 2016). In the expression of $G$ in equation (76), the scale factor ($a$) is an increasing function of time. For negative values of the parameter $n$, $G$ would be an increasing function of time. Therefore, in the present study we have used only negative values of the parameter $n$.

Using equations (21) and (76), we get,

$$\frac{\dot{G}}{G} = -n\frac{\dot{a}}{a} = -nH = -n\frac{\alpha\beta}{t_0}\left(\frac{t}{t_0}\right)^{\beta-1} \tag{77}$$

Writing equation (77) for the present time, i.e., $t = t_0$, one obtains,

$$n = -\frac{t_0}{\alpha\beta}\left(\frac{\dot{G}}{G}\right)_{t=t_0} \tag{78}$$

Using equation (78), the parameter $n$ can be estimated from the experimental observations regarding $\left[\dot{G}/G\right]_{t=t_0}$, as reported in articles on time varying gravitational constant (Ray, Mukhopadhyay and Dutta Choudhury 2007). According to a study by S. Weinberg, the largest possible value of $\left|\frac{\dot{G}}{G}\right|_{t=t_0}$ is $4 \times 10^{-10}$ per year (Weinberg 1972). Using equation (78), one gets $n \sim -5.5$ for this upper limit.

| TABLE – 2 |||||| 
|---|---|---|---|---|---|
| Values of $\omega_0, m, \gamma_0, \gamma_{D0}, \left(\frac{\dot{G}}{G}\right)_{t=t_0}$ for several values of the parameter $n$ which determines the time variation of the scalar field ($\phi$). ||||||
| $n$ | $\omega_0$ | $m$ | $\gamma_0$ | $\gamma_{D0}$ ($z_c = 0.7$) | $\left(\frac{\dot{G}}{G}\right)_{t=t_0}$ $(Yr^{-1})$ |
| -1.63 | -1.514 | -0.183 | -1.796 | -2.566 | $1.198 \times 10^{-10}$ |
| -1.64 | -1.518 | -0.168 | -1.691 | -2.416 | $1.205 \times 10^{-10}$ |
| -1.65 | -1.522 | -0.154 | -1.588 | -2.268 | $1.212 \times 10^{-10}$ |
| -1.66 | -1.525 | -0.140 | -1.486 | -2.123 | $1.220 \times 10^{-10}$ |
| -1.67 | -1.529 | -0.127 | -1.386 | -1.979 | $1.227 \times 10^{-10}$ |
| -1.68 | -1.532 | -0.115 | -1.287 | -1.839 | $1.234 \times 10^{-10}$ |
| -1.69 | -1.535 | -0.104 | -1.190 | -1.700 | $1.242 \times 10^{-10}$ |
| -1.70 | -1.538 | -0.093 | -1.095 | -1.564 | $1.249 \times 10^{-10}$ |
| -1.71 | -1.540 | -0.083 | -1.001 | -1.430 | $1.256 \times 10^{-10}$ |
| -1.72 | -1.543 | -0.073 | -0.909 | -1.299 | $1.264 \times 10^{-10}$ |
| -1.73 | -1.545 | -0.064 | -0.819 | -1.170 | $1.271 \times 10^{-10}$ |
| -1.74 | -1.547 | -0.055 | -0.730 | -1.043 | $1.279 \times 10^{-10}$ |
| -1.75 | -1.549 | -0.046 | -0.643 | -0.918 | $1.286 \times 10^{-10}$ |
| -1.76 | -1.551 | -0.038 | -0.558 | -0.796 | $1.293 \times 10^{-10}$ |
| -1.77 | -1.553 | -0.031 | -0.474 | -0.677 | $1.301 \times 10^{-10}$ |
| -1.78 | -1.554 | -0.024 | -0.392 | -0.559 | $1.308 \times 10^{-10}$ |
| -1.79 | -1.555 | -0.017 | -0.311 | -0.444 | $1.315 \times 10^{-10}$ |
| -1.80 | -1.557 | -0.011 | -0.232 | -0.332 | $1.323 \times 10^{-10}$ |
| -1.81 | -1.558 | -0.005 | -0.155 | -0.221 | $1.330 \times 10^{-10}$ |
| -1.82 | -1.559 | 0.001 | -0.079 | -0.113 | $1.337 \times 10^{-10}$ |
| -1.83 | -1.560 | 0.006 | -0.005 | -0.008 | $1.345 \times 10^{-10}$ |
| -1.84 | -1.561 | 0.012 | 0.067 | 0.095 | $1.352 \times 10^{-10}$ |
| -1.85 | -1.561 | 0.016 | 0.137 | 0.196 | $1.359 \times 10^{-10}$ |

Table-2 shows the values of $\omega_0$, $m$, $\gamma_0$, $\gamma_{D0}$, $\left(\frac{\dot{G}}{G}\right)_{t=t_0}$ for different values of $n$ which governs the time variation of the scalar field ($\phi$), as per equation (10). For the values of $n$ over the range of $-1.76 < n < -1.67$, the values of the present EoS parameter ($\gamma_0$) in this table are consistent with the range of values (approximately $-1.7 < \gamma_0 < -0.6$) obtained from astrophysical observations, as reported in some recent articles on the time variation of EoS parameter, based on an anisotropic space-time (Pradhan, Amirhashchi and Saha 2011; Yadav, Rahaman and Ray 2011). Over this range, the values of *m* are negative, indicating that *ω* would be less negative as $\phi$ increases, as evident from equation (11), where $\omega_0$ is negative. For $n < -1.81$ in this table we have $m > 0$, implying that *ω* would be more negative as $\phi$ increases. Both of these natures of variations of $\omega(\phi)$ are reflected in the plots of some other studies on Brans-Dicke theory (Chakraborty and Debnath 2009; Sharif and Waheed 2012). But the second type of variation is not physically valid, since it corresponds to those values of EoS parameter which are not in conformity with the astrophysical observations in this regard. The values of $\left(\frac{\dot{G}}{G}\right)_{t=t_0}$ in this table are absolutely within the range of permissible values shown by S. Weinberg (Weinberg 1972).

FIGURES

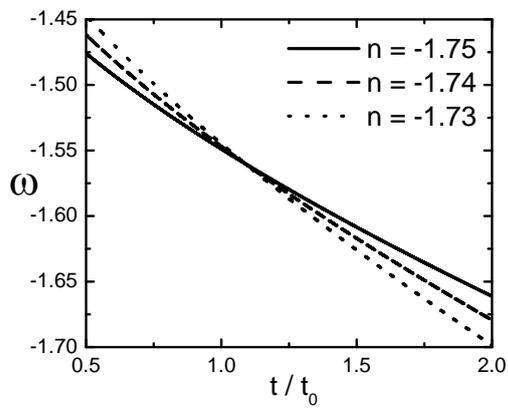

**Fig. 1** Plots of Brans-Dicke parameter versus time for three different values of the parameter $n$

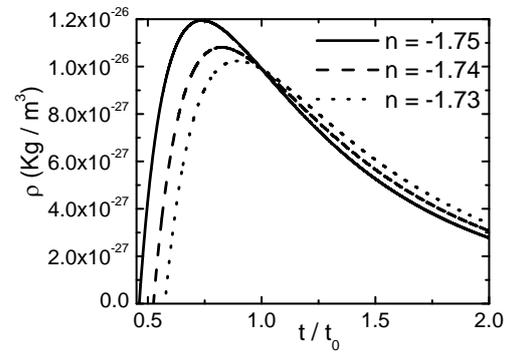

**Fig. 2** Plots of energy density versus time for three different values of the parameter $n$

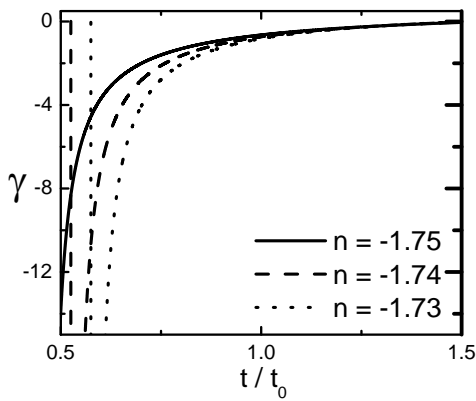

**Fig. 3** Plots of the equation of state parameter versus time for three different values of the parameter $n$

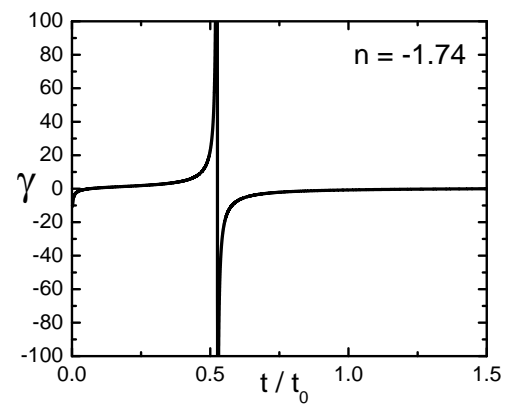

**Fig. 4** Plot of the equation of state parameter versus time, for $n = -1.74$, on a time scale longer than that of fig. 3

FIGURES

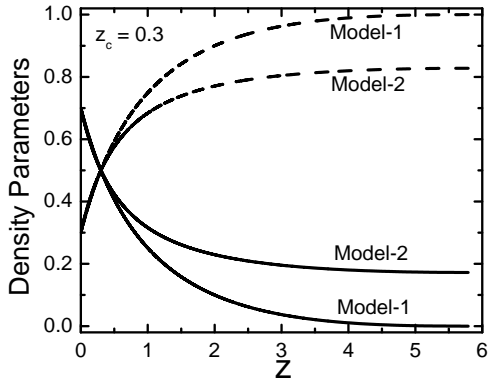

**Fig. 5** Plots of $\Omega_D$ (solid curves) and $\Omega_m$ (dashed curves), from two models, for $z_c = 0.3$, against the redshift parameter

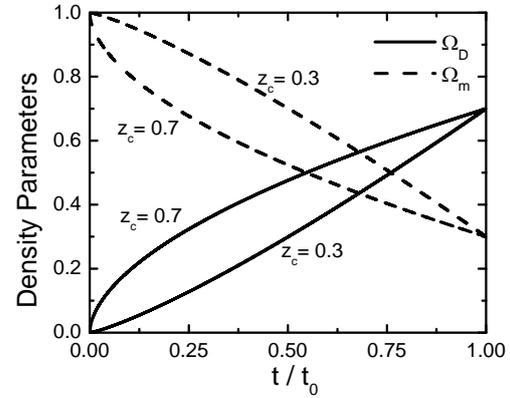

**Fig. 6** Plots of $\Omega_D$ and $\Omega_m$ (using model-1) as functions of time, for three different values of $z_c$

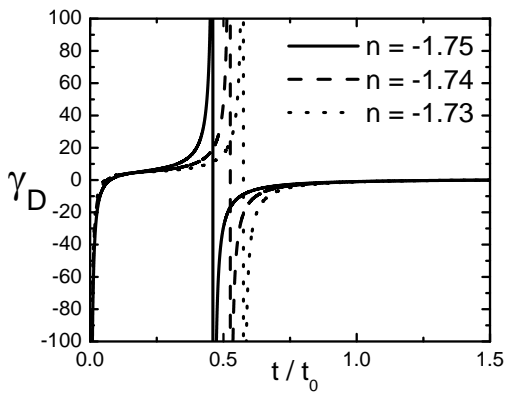

**Fig. 7** Plots of the equation of state (EoS) parameter for dark energy versus time for three different values of the parameter $n$

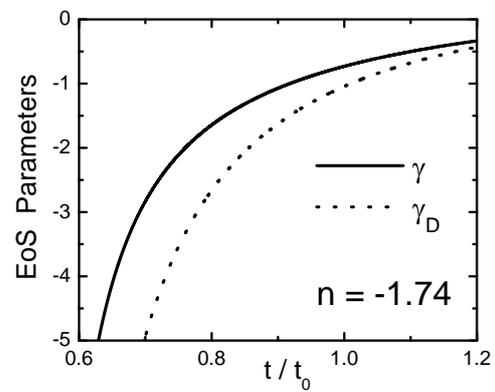

**Fig. 8** Plots of EoS parameters versus time, for total energy ($\gamma$) and dark energy ($\gamma_D$). For $n = -1.74$, $\gamma_0 = -0.73$ and $\gamma_{D0} = -1.04$

# FIGURES

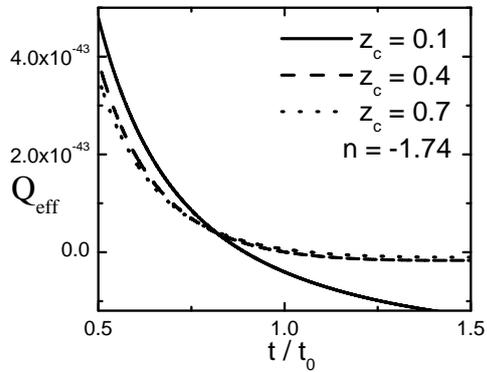

**Fig. 9** Plots of effective interaction term $Q_{eff}$ versus time for three different values of $z_c$

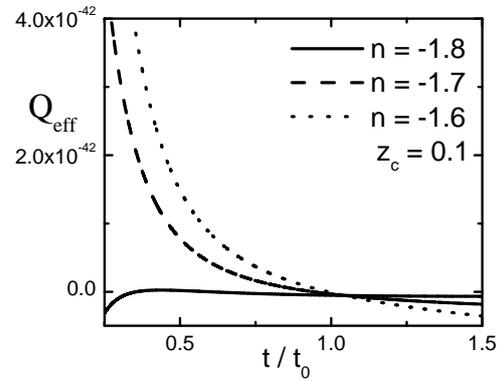

**Fig. 10** Plots of effective interaction term $Q_{eff}$ versus time for three different values of the parameter *n*

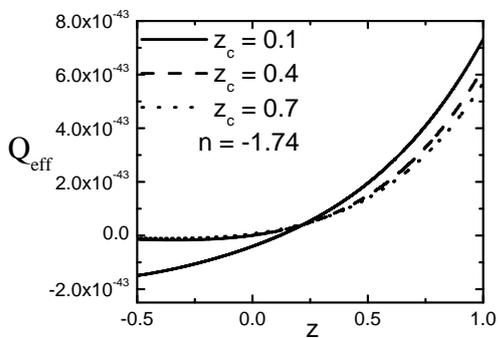

**Fig. 11** Plots of effective interaction term $Q_{eff}$ versus redshift (z) for three different values of $z_c$

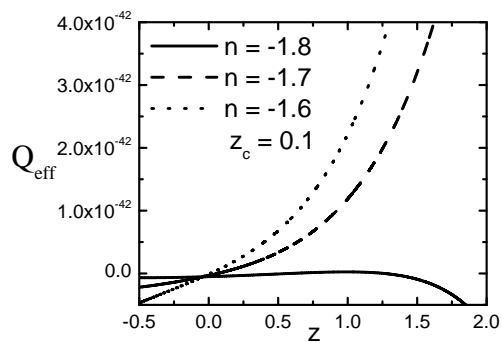

**Fig. 12** Plots of effective interaction term $Q_{eff}$ versus redshift (z) for three different values of the parameter *n*

FIGURES

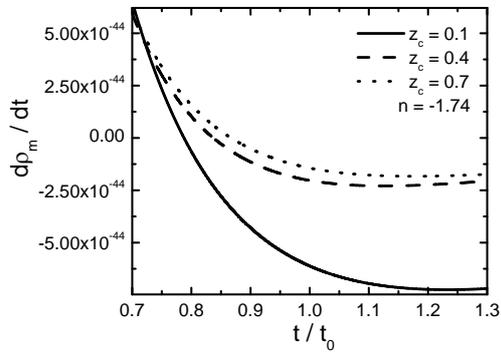

**Fig. 13** Plots of $\dot{\rho}_m$ versus time for three different values of $z_c$

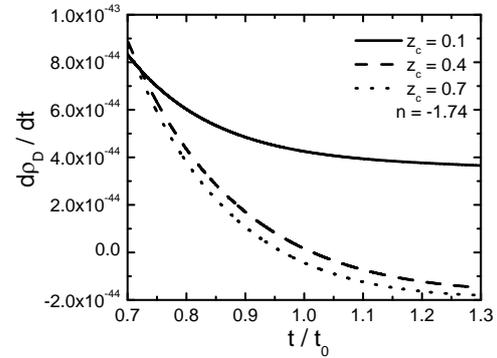

**Fig. 14** Plots of $\dot{\rho}_D$ versus time for three different values of $z_c$

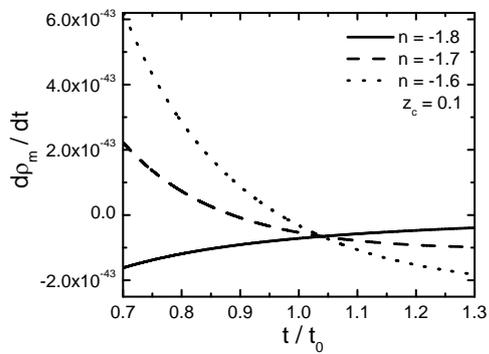

**Fig. 15** Plots of $\dot{\rho}_m$ versus time for three different values of the parameter $n$

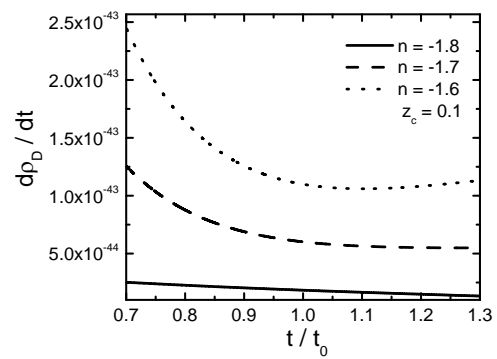

**Fig. 16** Plots of $\dot{\rho}_D$ versus time for three different values of the parameter $n$

## 7     Results and Discussion

Figure 1 shows the variation of the Brans-Dicke parameter ($\omega$) as a function of time, for different values of the parameter *n*. Here, $\omega$ has a negative value and it becomes more negative with time. For more negative values of *n*, it decreases less rapidly. This nature of $\omega$, which is gradually becoming more negative with time, is also obtained from other studies on Brans-Dicke theory (Chakraborty and Debnath 2009; Sharif and Waheed 2012). The time dependence of $\omega$, obtained from the present study, is based on a scale factor which leads to a time dependent deceleration parameter (that shows a signature flip with time from positive to negative), unlike some recent studies (Sahoo and Singh 2002; Jamil and Momeni 2011; Satish and Venkateswarlu 2014) that are based on time independent deceleration parameters. The value of $\omega_0$ (close to $-1.55$) and its gradual decrease with time are consistent with the results provided in the section of 'conclusions' of a recently published article based on Kantowski-Sachs space-time (Satish and Venkateswarlu 2014).

Figure 2 shows the time variation of energy density ($\rho$) with time, for different values of the parameter *n*. Here, $\rho$ rises steeply to a peak and then decreases at a slower rate. More negative values of *n* cause quicker attainment of the peak and also a larger peak value of energy density. It falls more steeply for more negative values of *n*. This behaviour is consistent with studies based on General Relativity (Pradhan, Amirhashchi and Saha 2011).

Figure 3 shows the time dependence of the equation of state (EoS) parameter ($\gamma$) for three different values of the parameter *n*. These curves show a steep rise initially and then they gradually become asymptotic to a small negative value. Saturation is reached faster for less negative values of *n*. The value of $\gamma$ at the present time ($t = t_0$) is $-0.73$. The ranges of $\gamma$ from observational results based on SN Ia data and SN Ia data collaborated with galaxy clustering statistics and CMBR anisotropy are $-1.67 < \gamma < -0.62$ and $-1.33 < \gamma < -0.79$ respectively. (Knop et al. 2003; Tegmark et al. 2004). These curves are similar to those obtained from dark energy models based on LRS Bianchi type-V metric in the framework of Einstein's general theory of relativity (Yadav, Rahaman and Ray 2011).

Figure 4 shows a single plot EoS parameter, for $n = -1.74$, on a longer time scale. It shows a pattern that is very much similar to the one depicted in one of the earlier studies based on dark energy models in anisotropic Bianchi type-I (B-I) space-time, in the framework of Einstein's general theory of relativity (Pradhan, Amirhashchi and Saha 2011).

The plots in the figures 2-4 show very clearly that the time evolution of energy density and EoS parameter, obtained here for FRW space-time in the framework of Brans-Dicke theory, are quite in agreement with the results obtained from anisotropic space-time in the framework of general relativity.

Figure 5 shows the variations of density parameters, for dark energy and matter, as functions of the redshift parameter, using models-1, 2 (Sec. 4.1, 4.2). Here, the value of $\mu$ has been taken to be $-0.557$, causing $z_c = 0.7$, implying that the dark energy took over at around $z = 0.7$. These plots are quite similar in nature to the plots obtained from some recent studies (Das and Al Mamon 2014; Paul, Thakur and Ghose 2010) which were carried out on premises completely different from the present study.

Figure 6 shows the plots of density parameters, for dark energy and matter, as functions of time (Sec. 4.1), for three different values of $z_c$. For higher values of $z_c$ (implying less negative values of $\mu$, as per eqn. 39), $|\dot{\Omega}|$ gradually decreases with time.

Figure 7 shows the time variation of the EoS parameter for dark energy ($\gamma_D$) for three different values of the parameter *n*. These plots are very much similar to the ones obtained from studies based on anisotropic space-time in the framework of general relativity (Pradhan, Amirhashchi and Saha 2011; Yadav, Rahaman and Ray 2011). Plots with less negative values of *n* take longer time to attain saturation.

Figure 8 shows the time variations of EoS parameters for total energy and dark energy. Here $\gamma_D$ has a more negative value compared with $\gamma$. Their difference decreases with time, as the proportion of dark energy increases in the universe. For $n = -1.74$ the values of $\gamma_0$ and $\gamma_{D0}$ are respectively $-0.73$ and $-1.04$ respectively. These values are quite consistent with the ranges of values obtained from several astrophysical observations and described in some studies regarding dark energy and time varying EoS parameter (Pradhan, Amirhashchi and Saha 2011; Yadav, Rahaman and Ray 2011; Das and Al Mamon 2014).

The parameter *n*, in this model, plays an important role in governing the time evolution of various cosmological quantities. Both $\omega_0$ and *m*, which controls the value of $\omega$, are dependent upon *n* (eqns. 16, 17). Larger negative value of *n* means faster change of $\phi$ ($\equiv 1/G$) with time. Taking $n = -1.74$ we get, $\left(\frac{\dot{G}}{G}\right)_{t=t_0} = 1.28 \times 10^{-10} \, Yr^{-1}$ from equation (77). This is well within its upper limit, i.e. $4 \times 10^{-10} Yr^{-1}$, predicted by S. Weinberg (Weinberg 1972).

Figure 9 shows the variation of the effective interaction term $Q_{eff}$ as a function of time for three different values of $z_c$. It is found to decrease with time from an initial positive value, gradually becoming negative at a later time. This behaviour of signature flip is consistent with the results of a recent study on holographic dark energy (Zadeh, Sheykhi and Moradpour 2017). For smaller values of $z_c$ (which correspond to more negative values of $\mu$), $Q_{eff}$ attains its negative phase faster. Its positive phase indicates that dark energy initially decayed into matter and, in the later stage, where $Q_{eff} < 0$, matter decays into dark energy.

Figure 10 shows the variation of $Q_{eff}$ as a function of time for three different values of the parameter *n*. Two of these curves decrease with time and change their signs from positive to negative. These two curves show that the transfer of energy, between matter and dark energy, reverses its direction, producing dark energy at the expense of matter at the later stage. The third curve, with $n = -1.8$, is entirely in the negative domain, becoming gradually closer to zero with time. It clearly shows that, there exists a value of the parameter *n*, below which we have an unidirectional energy transfer, from the matter field to the field of dark energy, at a gradually decreasing rate.

Figure 11 depicts the change of $Q_{eff}$ as a function of z, for three different values of $z_c$. It is found that, for smaller values of $z_c$, $Q_{eff}$ attains its negative phase faster.

Figure 12 shows the variation of $Q_{eff}$ as a function of the redshift parameter (z) for three different values of the parameter $n$. Like figure 10, one of these three plots also indicate the existence of a threshold value for $n$, below which the direction of energy transfer is always from the matter field to the field of dark energy with $Q_{eff}$ becoming less negative with time. Above this threshold of $n$, the interaction term shows a signature flip from positive to negative as time progresses. The curve for $n = -1.8$ in this figure is similar to the behaviour obtained from a study on the interaction term between dark matter and dark energy (Cueva and Nucamendi 2010). Since around 80% of the matter content is dark matter, the interactions of dark energy with matter consist mainly of the interactions of dark energy with dark matter.

Figure 13 shows the variation of $\dot{\rho}_m$ as a function of time for three different values of $z_c$. Each graph shows a decrease with time, becoming positive to negative prior to the present time ($t = t_0$). Smaller values of $z_c$ cause faster transition from positive to negative values. In an interacting model of matter and dark energy, a negative value of $\dot{\rho}_m$ means a transfer of energy from the sector of matter to that of dark energy.

Figure 14 depicts the variation of $\dot{\rho}_D$ as a function of time for three different values of $z_c$. Each plot shows a decrease of $\dot{\rho}_D$ with time. For the larger values of $z_c$, we find a faster fall of these graphs. For a conversion matter into dark energy, at the present time, it is necessary that $\dot{\rho}_D$ is positive when $\dot{\rho}_m$ is negative. Figures 13, 14 show that it is possible to find a value of $z_c$ for which this condition is satisfied during a span close to the present time ($t = t_0$).

Figure 15: shows the variation of $\dot{\rho}_m$ as a function of time for three different values of the parameter $n$. For less negative values of this parameter, these curves show a decrease with time, from positive to the negative region. For more negative values of $n$, $\dot{\rho}_m$ remains negative throughout, with a gradually decreasing slope. For less negative values of $n$, $\dot{\rho}_m$ becomes negative prior to the present time ($t = t_0$). It means that the matter content started decaying into dark energy somewhere in the recent past, causing an accelerated expansion of the universe.

Figure 16 shows the time variation of $\dot{\rho}_D$ for three different values of the parameter $n$. For $z_c = 0.1$, $\dot{\rho}_D$ is positive for all $n$ values here (implying a creation of dark energy) and it decreases with time. The curves with more negative values of $n$ are lying below the ones with less negative values of the same parameter. It is evident that the more negative values of $n$ cause slower rate of production of dark energy.

## 8    Concluding Remarks

The present study is based on a spatially flat, homogeneous and isotropic space-time, in the framework of Brans-Dicke theory. Here, the time evolution of Brans-Dicke parameter ($\omega$), energy density ($\rho$) and EoS parameter ($\gamma$) have been determined from the field equations. The choice of the scale factor, that ensures a signature flip of the deceleration parameter with time, is likely to give more credibility to the results of this model. Without involving any self interaction potential $(V_\phi)$ and cosmological constant ($\Lambda$), the time evolutions of density parameters for matter and dark energy have been determined and these results are sufficiently consistent with those obtained through different models and methods (Das and Al Mamon 2014; Paul, Thakur and Ghose 2010). The parameters $\mu$ and $\lambda$, in model-1 and model-2 (sec. 4.1, 4.2), govern the time dependence of $\Omega_D$, $\Omega_m$ and $\gamma_D$. It has been shown that $\mu$ and $\lambda$ can be determined if one gets information from astrophysical observations regarding the time or redshift ($z$) at which we had $\Omega_D = \Omega_m$ in the recent past of the expanding universe. Assuming an interaction to be taking place between matter and dark energy, an effective interaction term has been calculated on the basis of the energy conservation equation. For some combinations of $n$ and $z_c$ (or $n$ and $\mu$), this term changes its sign from positive to negative, showing a change of direction of energy transfer between the sectors of matter and dark energy. Its negative value at the present time implies a decay of matter into dark energy which is responsible for the accelerated expansion of the universe. The behaviour of this interaction term is consistent with the results of some recent studies (Zadeh, Sheykhi and Moradpour 2017; Cueva and Nucamendi 2010) carried out in a completely different way. Using the expression of $Q_{eff}$ we have determined the time variations of $\dot{\rho}_m$ and $\dot{\rho}_D$. The conditions under which they respectively have negative and positive signs during a span of cosmic time, matter may be regarded as decaying into dark energy in that phase of expansion of the universe. The parameter $n$, which actually controls the rate of change of gravitational constant with time (as per eqn. 76), determines the nature of time variation of some cosmological quantities ($\omega$, $\rho$ and $\gamma$) mentioned above. It implies that the dependence of the scalar field ($\phi \equiv 1/G$) upon the scale factor ($a$) (eqn. 10) plays a very important role in cosmic expansion. The dependence of $\omega$ upon $\varphi$ is controlled by $\omega_0$ and $m$ which are all functions of $n$. Positive values of $n$ leads to large negative values of $\gamma_{D0}$, contrary to its range of values close to $-1$, obtained from recent astrophysical observations (Das and Al Mamon 2014). This result is in favour of using negative values of $n$, implying an increase of gravitational constant with time, as per equation (76). This behaviour of gravitational constant is also obtained from other studies (Roy, Chattopadhyay and Pasqua 2013; Pradhan, Saha and Rikhvitsky 2015). Keeping in view all these facts, one may think of improving this model by choosing a relation, between $\phi$ and $a$, which is different from the ansatz of equation (10), in order to have more than one tunable parameter like $n$ controlling the time variation of the scalar field ($\phi$). As a future project, one may think of determining the time dependence of $\rho_m$ and $\rho_D$ from equations (59) and (60) by assuming $Q$ to have a form like $Q = \kappa H$ where $\kappa$ is a constant and $H$ is the Hubble parameter. In the second equation, $\gamma_D$ can be replaced by $\gamma_{D0}$, considering a very slow time variation of $\gamma_D$, as evident from figure 8, over a span of time close to $t = t_0$. The time dependence of $H$ can be taken from equation (21). Using the solutions of these differential equations, (59) and (60), one can then calculate $\Omega_D$ and $\Omega_m$, taking the energy density ($\rho$) from equation (15). It would be possible to make an estimate of $\kappa$ by using the known values of $\Omega_{D0}$ and $\Omega_{m0}$ at $t = t_0$ in the expressions of the density parameters for matter and dark energy. Using equations (10) and (19), one would be able to express the density parameters $\Omega_D$ and $\Omega_m$ in terms of the scalar field ($\phi$).

## Acknowledgements

The authors of this article express their thanks and gratitude, very heartily, to the researchers and academicians whose works have enriched them immensely and inspired them to undertake the present study. S. Roy would like to thank all his colleagues in the Department of Physics, St. Xavier's College, Kolkata, India for creating and maintaining an ambience of erudition and an atmosphere of cooperation and support. The authors also thank the anonymous reviewers of this article for their valuable comments and useful suggestions.